# Towards manipulating relativistic laser pulses with 3D printed materials

L. L. Ji[1], J. Snyder[1], A. Pukhov[2], R. R. Freeman[1] & K. U. Akli[1]

**Efficient coupling of intense laser pulses to solid-density matter is critical to many applications including ion acceleration for cancer therapy. At relativistic intensities, the focus has been mainly on investigating various laser beams irradiating initially flat interfaces with little or no control over the interaction. Here, we propose a novel approach that leverages recent advancements in 3D direct laser writing (DLW) of materials and high contrast lasers to manipulate the laser-matter interactions on the micro-scales. We demonstrate, via simulations, that usable intensities $\geq 10^{23}$Wcm$^{-2}$ could be achieved with current tabletop lasers coupled to 3D printed plasma lenses. We show that these plasma optical elements act not only as a lens to focus laser light, but also as an electromagnetic guide for secondary particle beams. These results open new paths to engineering light-matter interactions at ultra-relativistic intensities.**

[1]Physics Department, The Ohio State University, Columbus, OH 43210, USA. [2]Institut für Theoretische Physik I, Heinrich-Heine-Universität Düsseldorf, 40225 Düsseldorf, Germany. Correspondence and requests for materials should be addressed to L. L. J. (email: ji.289@osu.edu) or K. U. A. (emial: akli.1@osu.edu).

The manipulation of laser light has led to many spectacular advances in medicine, telecommunication[1], remote sensing[2], and in probing the fundamental interaction of light with matter at the atomic scales[3]. Controlling and manipulating laser light at relatively low intensities has been widely studied using photonic crystals[4], heterostructures[5], and metamaterials[6]. As the laser intensity has increased beyond the damage threshold, these optical approaches are less effective as the laser pulse efficiently ionizes the materials. Further intensity increase moves the interaction into the relativistic regime[7] ($>10^{18}$Wcm$^{-2}$), where many exotic phenomena have been predicted and experimentally observed in laser-matter interactions. These include the production of relativistic electrons[8-14], the acceleration of protons and heavy ions[15-22], the synthesis of attosecond pulses from plasma-induced harmonics[23-26], and the creation of electron-positron jets[27]. The investigation of these processes has been focused on exploring their dependence on various laser pulses as well as target parameters (spatial dimensions, density, and atomic number). More recently, front surface target morphology is being introduced to enhance laser matter interactions. Structured interfaces including sub-wavelength structures, snowflakes, nanowires and cone targets are used to enhance laser absorption[29-32], to guide the electrons/protons[33-35], or to serve as plasma undulators[36].

Despite recent progress, controlling light-matter interactions at relativistic intensities has not been extensively explored for two main reasons. First, the manufacturing of advanced micro- and nano-structures has been the domain of other scientific disciplines and their use is confined to intensities below the damage threshold. Second, high-intensity short-pulses are inherently preceded by nanosecond scale pedestals[37] that can destroy or substantially modify any guiding structure due to hydrodynamic expansion. However, with recent advancement in laser technology and laser pulse cleaning techniques it is now possible to substantially reduce the undesired pedestals. Laser to pedestal contrast ratios higher than $10^{10}$ have been achieved using cross-polarized wave generation (XPW) technique[38]. Furthermore, highly repeatable structures with features as small as 100 nm can be easily manufactured using 3D direct laser-writing (DLW)[39]. The small size of the light guiding features and the short duration of the current generation of high contrast laser (tens of femtoseconds) make it possible to optimize the interaction using realistic and computationally manageable particle-in-cell (PIC) simulations.

We show via PIC simulations that, one of the most critical parameters-the laser intensity, can be manipulated using 3D printed hollow micro-cylinders. The structured target exhibits the behavior of a resonating optic. When a high-contrast intense laser pulse is input into the Micro-Tube Plasma (MTP) lens, it is significantly intensified. The new technique is able to boost light intensity up to $10^{23}$Wcm$^{-2}$ based on today's high power laser systems, allowing for the exploration in the exotic near-QED (quantum electro-dynamics) regime. Further, by manipulating the laser intensity, controlling or maximizing the outcome of laser-plasma interaction becomes possible. We show that the interaction of an intense laser beam with flat target via an MTP lens can increase the on-target laser intensity by concentrating the energy in a smaller focal spot. As a result, the generation of secondary electron, proton, and γ-ray beams are enhanced.

## Results

**Light intensification in a micro-tube plasma target.** The proposed scheme is shown in Fig. 1a. The target consists of periodic hollow micro-cylinders attached to a flat substrate. These guides can now be printed using two-photon lithography process available on most commercial 3D printers with a featured resolution around 200 nm. With many identical micro-tubes, the laser will enter one of them with a good and certain opportunity. To get insight into the interaction of intense light with micro-tubes, we have carried out 3D PIC simulations of a high-contrast laser pulse with a single tube (Fig.1b). A laser beam with a duration of 40 fs and intensity of $5.3 \times 10^{21}$ $Wcm^{-2}$ enters the target via a micro-tube of 4.8 μm diameter. Comparing the beam intensities outside and inside the tube reveals substantial enhancement. The laser intensity is boosted by a factor of ~3 inside the tube. A peak intensity of $1.5 \times 10^{22}$ Wcm$^{-2}$ is reached after the pulse has propagated a distance of 8 µm from the entrance aperture. Increasing laser beam intensity requires increasing the total laser energy, shortening the pulse duration, and/or decreasing the laser focal spot. Detailed analyses of the spatial and temporal characteristics of the laser indicate that the light intensification is due to tight focusing of the beam. The left and right plots in Fig.1c show the laser spatial distributions in the transverse $y-z$ plane averaged along the propagation axis $x$. The dashed white line indicates the tube boundary. It is evident from these spatial profiles that the micro-tube acts as a focusing optical element. A tight focal spot with a full-width-half-maximum (FWHM) of $1.8 \mu m$ is achieved inside the tube compared to $2.9 \mu m$ (FWHM) for the input beam. Temporal snapshots of the beam did not reveal any pulse duration shortening (Fig.1c). Both input and in-tube laser pulses show similar oscillatory motions in time.

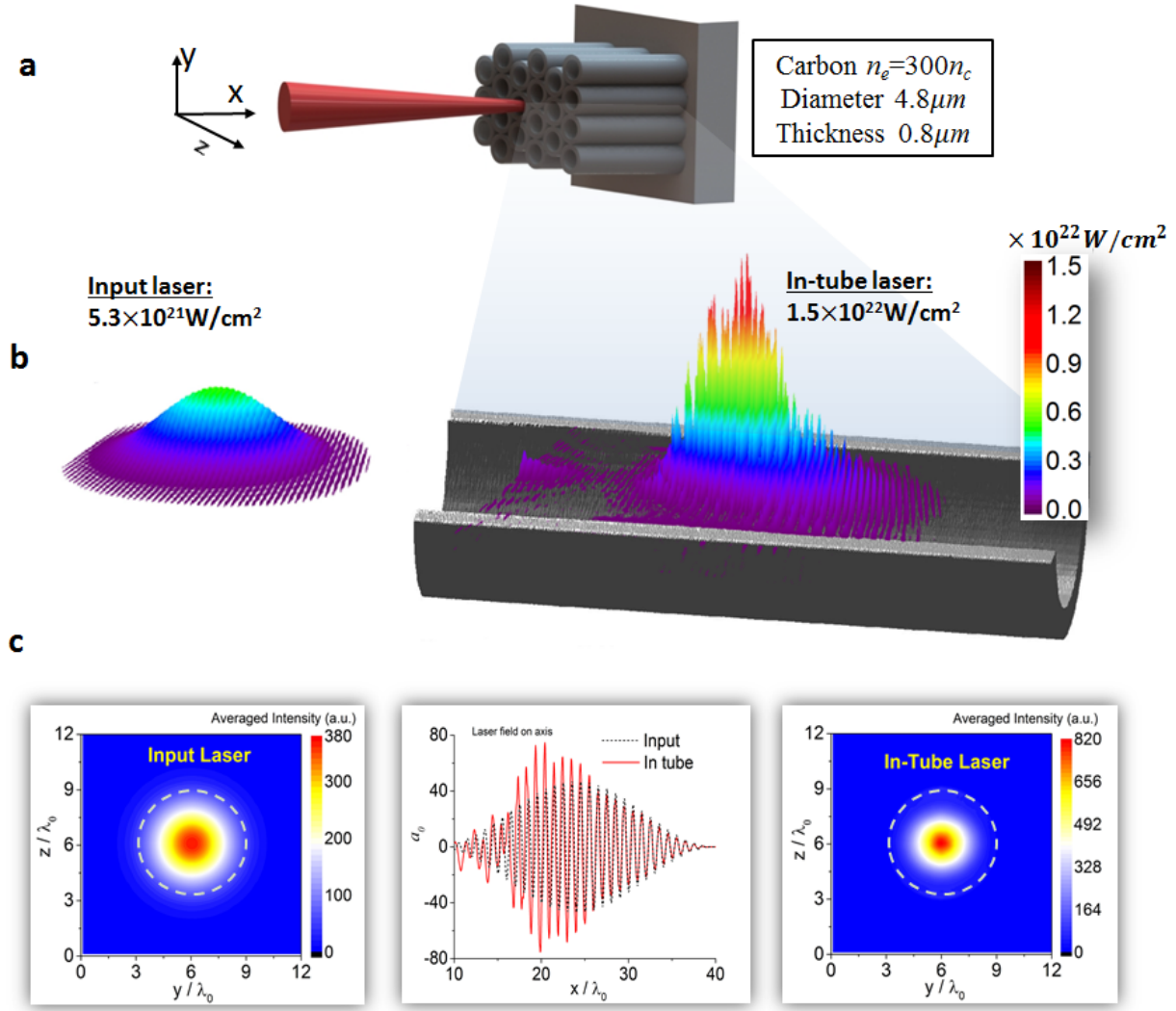

**Figure 1 | Scheme of laser-micro-tube interaction.** (**a**) Design of a relativistic fs laser impinging on a periodic micro-tube target. (**b**) Iso-surface plots for the laser intensity distribution before and after it enters the tube. (**c**) Light intensity distribution on the *x-y* plane for the input pulse and in-tube pulse.

**Mechanism and scaling of light intensification**

We performed a series of 3D PIC simulations to systematically study light intensification in a hollow micro-cylinder with a fixed diameter (4.8 µm). The laser intensity was varied from $10^{17}$ Wcm$^{-2}$ to $10^{23}$ Wcm$^{-2}$. The peak laser intensity inside the tube at each simulation time is recorded for all input pulses. The light intensification factor is determined as the ratio of the in-tube to the input pulse intensity ($\eta = I_{in-tube} / I_{input}$). This quantity is plotted in Fig. 2a as a function of the input laser intensity. Three distinct regimes of intensification are clearly observed.

<u>Diffraction regime</u>: In this regime, the light intensification is independent on the laser intensity. As the electron density distribution in Fig. 2b indicates, there is no background plasma inside the tube. The interaction of the laser pulse with the tube is predominantly determined by the size of the aperture. As a result the narrowing of the focal spot is mainly due to diffraction. The intensification happens in the near-field Fresnel region, since the aperture size is large compared to the light wavelength. The intensification factor is constant at around $\eta \approx 2.6$ for input laser amplitudes below $I_{input} \sim 10^{19}$ Wcm$^{-2}$.

Depletion regime: In the second regime, the laser field is strong enough to pull-off a considerable portion of electrons from the wall of the micro-tube. These laser-induced electrons form a low density cloud, below one tenth of the critical density (Fig. 2c). These electrons are distributed with fluctuation and disorder (see also in Fig. 2e). In addition to being diffracted, the laser field is depleted by the under-dense plasma electrons, leading to a slightly lower intensification factors compared to the ones in the diffraction regime.

Focusing regime: For laser intensities higher than $I \approx 6 \times 10^{21}$ Wcm$^{-2}$, we observe a monotonic increase of the in-tube pulse intensity. In this domain of LPI, the laser amplitude is high enough to pull-off a substantial fraction of the electrons from the walls of the micro-tube. These electrons stay together, forming periodic dense bunches, as shown in Fig. 2d and f. The averaged electron density in the tube is shown in fig. 2e. At the inner boundary, in the vicinity of the wall, the formed plasma is over-dense (i.e. with density equal or higher to the critical density). Within this laser-induced over-dense layer, the outer edge of the laser field decays rapidly to zero in a skin depth. This layer changes the tube geometry by acting as a new tube boundary. The laser pulse sees a diameter-reduced tube in this case and the laser energy is confined to smaller focal spot.

The plasma density is lower on axis and higher when approaching the boundary. The V-shaped plasma density profile also contributes to the focusing for the propagation laser pulse. The intensification factor increases with input intensity. A laser beam with an initial intensity of $2.4 \times 10^{22}$ Wcm$^{-2}$ could be boosted to a peak intensity $10^{23}$ Wcm$^{-2}$ using the MTP device. The experimentally unexplored near-QED (quantum electro-dynamics) regime could be within reach on the existing laser facilities using micro-tube plasma devices coupled to primary targets.

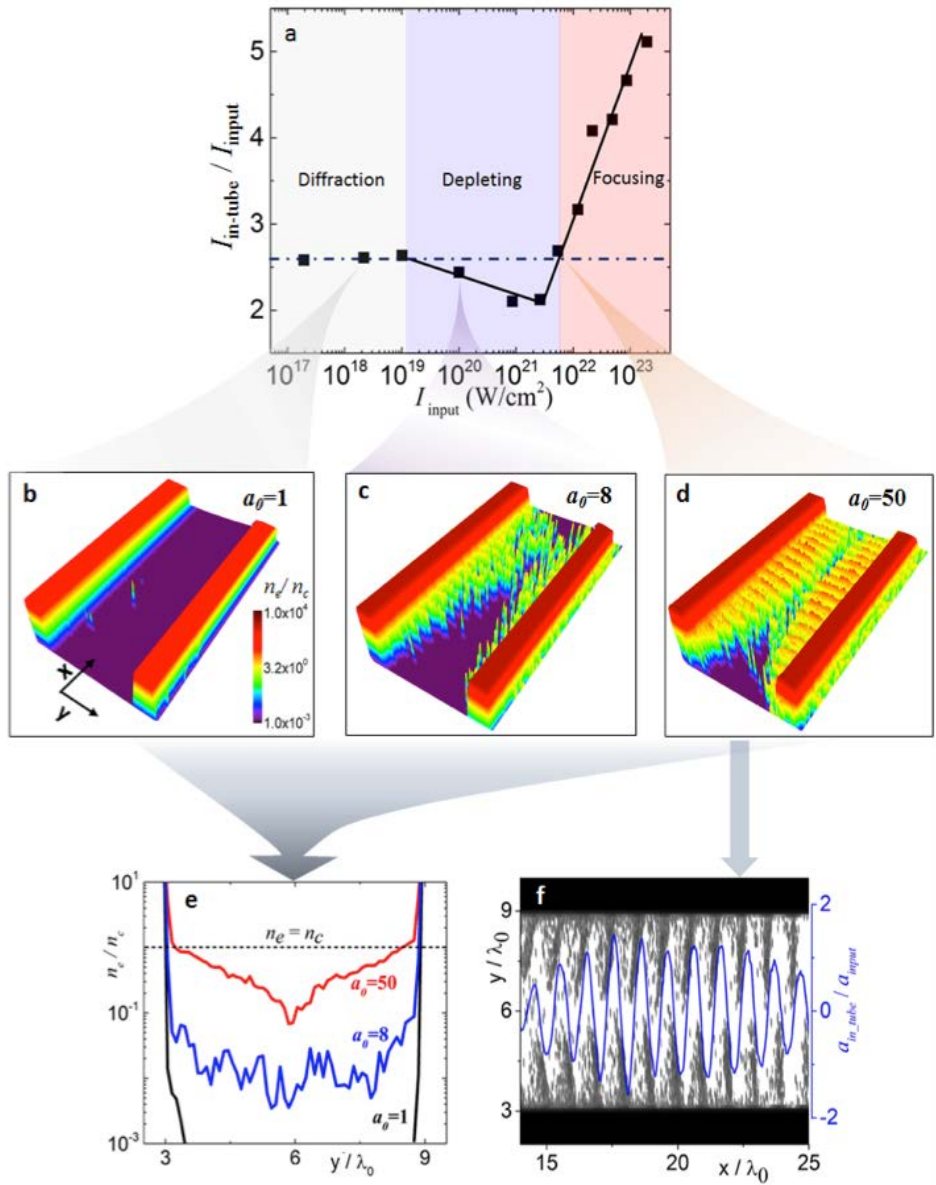

**Figure 2 | Scaling of light intensification as a function of light intensity.** (**a**) The light intensification factor in three difference intensity regions. (**b**)-(**d**) Typical electron density distribution in the diffraction, depleting and focusing regions, respectively. (**e**) Averaged electron density lineouts along the transverse axis *y*. (**f**) Laser field and electron density in the *x-y* plane for $a_0 = 50$.

**Thresholds between three regimes:** The transition from the diffraction to the depletion regime occurs when the laser pulse starts to create plasma in the tube. As electrons are dragged out of the tube by the transverse laser field, they are also accelerated forward directly by the Lorentz force, specifically by direct laser acceleration (DLA) mechanism. Efficient DLA requires that an electron is accelerated to a relativistic velocity in half laser period. If the velocity of an electron is much smaller than the speed of light, it will slip into the decelerating phase in less than half of the laser period, where the direction of the laser field is opposite to the one in the accelerating phase. The electron is quickly driven outwards and there is no background plasma in the tube. It is when the laser field at the tube boundary becomes relativistic that the electrons can be detached and stay in the tube. This gives

$$a_{r=r_0} \geq 1. \tag{1}$$

Here $a = eE_L/m_e\omega_0 c$ is the dimensionless laser electric amplitude and $r = \sqrt{\Delta y^2 + \Delta z^2}$ is the vertical distance from the propagation axis (the inner tube boundary is located at $r_0$). For the pulse profile and tube size we employed, Eq. 1 produces $a_0 \geq e$ ($e \approx 2.72$, the Euler number), corresponding to a peak intensity of $I_0 \approx 1.6 \times 10^{19}\,\text{Wcm}^{-2}$, in good agreement with Fig. 2a.

The threshold between the depleting and focusing regimes corresponds to when the electron density near the inner wall becomes over-dense, i.e., $\bar{n}_e/n_c \geq 1$ at $r = r_0$. When an electron is detached from the tube, it undergoes the transverse electric force from both the laser field and the charge separation field. The laser pulse cannot drag more electrons out as its transverse field is balanced by the charge separation one, $\bar{E}_y \sim \bar{E}_r$ at $r = r_0$. From Fig. 2e, we assume the plasma density drops linearly to zero from the boundary to the axis $\bar{n}_e(r) = (r/r_0)\bar{n}_{r_0}$. The charge separation field, at the inner boundary, is then $\bar{E}_{r_0} \sim 4\pi e\bar{n}_{r_0}r_0/3$. For the averaged laser field, as shown in Fig. 2f, the part $E_y > 0$ contributes to the electrons from the upper boundary ($y > 6\lambda_0$) and vice versa. The effective laser field averaged in one laser period is $\langle |E_{L,r_0} \sin(\omega_0 t)| \rangle / 2$. For the laser profile and tube radius we use in simulations, the dimensionless laser amplitude can be written as $\bar{a}_{r_0} = a_0 / e\pi$. The electron density at the boundary is obtained by balancing the two fields. This gives,

$$a_{thr} \approx \frac{2\pi^2 e}{3}\left(\frac{r_0}{\lambda_0}\right)\left(\frac{\bar{n}_{r_0}}{n_c}\right), (e \approx 2.72) \qquad (2)$$

The laser amplitude required to create critical plasma density is $a_{thr} \approx 53$ ($I_0 \approx 6 \times 10^{21} Wcm^{-2}$) for $r_0 = 3\lambda_0$, which is consistent with the laser amplitude observed in Fig. 2a.

**Applications of the MTP target:** The fact that the laser beam maintains its space-time integrity and is enhanced in intensity suggest that 3D printed structures could be used, as an intermediary micro-optical element, to increase the intensity on a flat foil at the end of the tunnel and thus manipulate or maximize the outcome of the LPI. As a case study, we carried out an investigation using an input laser intensity of $5 \times 10^{22} Wcm^{-2}$ ($a_0 = 150$), which lies in the focusing regime. We compare two cases: 1) laser beam interacting with a traditional carbon-hydrogen (CH) flat target (Fig .3a), 2) laser beam interacting with an identical flat target coupled to a $8\mu m$ long 3D printed micro-tube (Fig. 3b). The introduction of the intermediary micro-optical plasma elements increases the peak intensity on the secondary flat interface to $4.3 \times 10^{23} Wcm^{-2}$. We observe significant enhancements in terms of maximum and total energy of laser-induced electron, proton, and $\gamma$-ray beams when the micro-tube plasma lens is used (Fig. 3.c-e). Due to MTP lens, the maximum electron beam energy is increased from ~200 MeV to ~400 MeV. The laser-to-electron energy conversion efficiency for high-energy electrons (>50MeV) is enhanced by a factor of 8 in the compound target. Consequently, the $\gamma$-ray beam production is enhanced by a similar factor as the high-energy electrons are decelerated by the radiation back-reaction. With a micro-tube plasma lens, the maximum photon energy reaches 50 MeV compared to only 20 MeV with a traditional

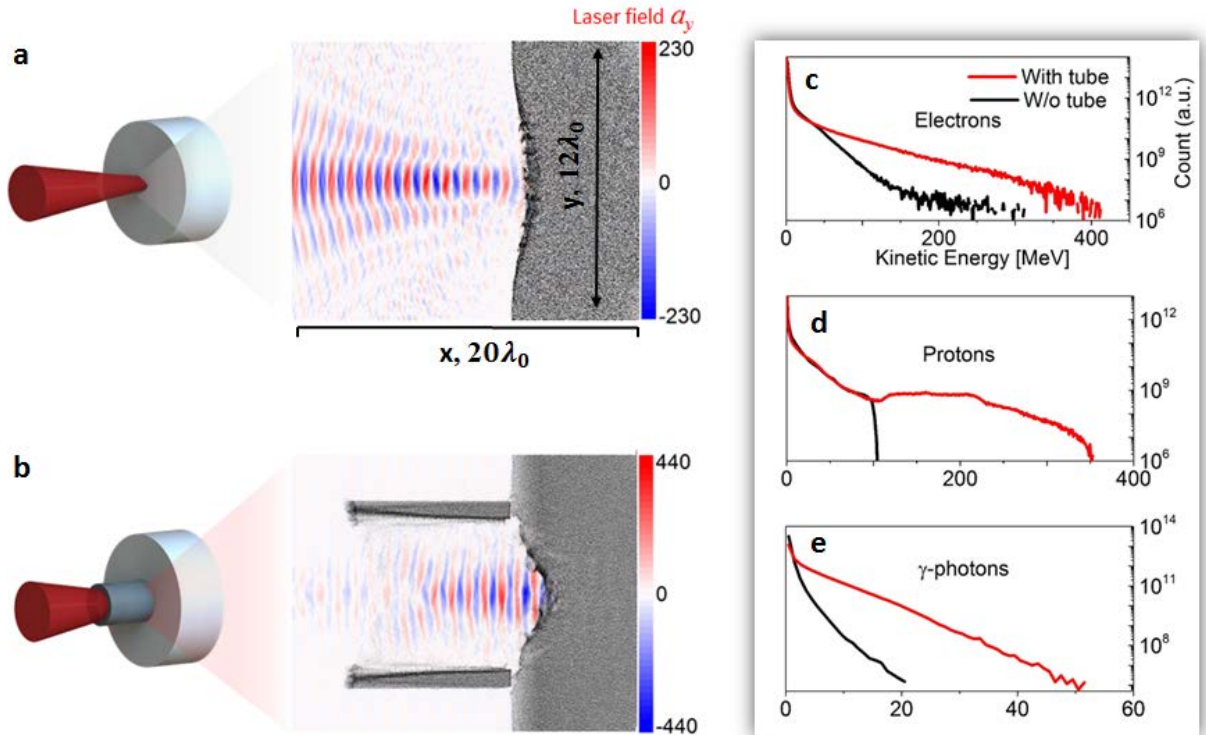

**Figure 3 | Simulation results of using flat target and MTP target**. (**a**) and (**b**) are the target setup and the corresponding distributions of the laser field and electron density. (**c**), (**d**) and (**e**) shows the spectrum comparison for electrons, protons and gamma-photons at 160fs, respectively

flat target. The foil at the end of the micro-tube is strongly deformed due to the increased laser pressure of the intensified laser pulse. The traditional target is only slightly bent, indicating a much slower shock wave due to the weaker on-target laser intensity. The maximum proton energy obtained with compound targets is ~350 MeV compared to only 100 MeV for flat targets for the same input laser intensity. The total number of energetic protons increases fivefold with light intensification.

## Discussion

The last 30 years has seen the development of ever increasing peak intensity from pulsed lasers. The quest for extremely intense laser pulses is largely motivated by the possibility of reaching experimentally unexplored regimes of light-matter interactions as well as the production of intense ion beams for cancer therapy. To increase the light intensity, one or a combination of the pulse parameters (energy, duration, and spot size) must be manipulated accordingly. Increasing the energy and/or shortening the pulse require using large optical elements due to damage threshold. Other nonconventional approaches such as self-focusing[40], coherent focusing of harmonics[41], and flying mirror[42] have been proposed. Here, we have outlined a novel approach that leverages recent advances in 3D printing of materials and high contrast lasers to boost light intensity using hollow micro-cylinders. The use of the 3D printed micro-optical elements provides another degree of freedom that makes it possible to micro-engineer laser plasma interactions. By controlling and adjusting the spatial dimensions (aperture size, length, thickness,..) of the 3D printed MTP targets, the pulse intensity can be manipulated. Hence various LPI applications can be tuned for optimal performance. These new results will open new paths towards micro-engineering laser plasma interactions that will benefit high field science, laser-based proton therapy, laser and particle beams, near-QED physics, nuclear physics, and relativistic nonlinear optics.

## Methods

**Numerical modeling of the laser interaction with micro-tube plasma target.** We simulate laser-plasma interaction with the full-3D particle-in-cell code VLPL[43]. In a simulation box of $40\lambda_0 \times 12\lambda_0 \times 12\lambda_0$ ($\lambda_0 = 0.8 \mu m$ is the laser wavelength) in $x \times y \times z$ directions, a laser pulse polarized in the $y$ direction enters from the left boundary along the $x$ direction. The laser field amplitude has a profile of $a_y = a_0 e^{-r^2/\sigma_0^2} \sin^2(\frac{\pi t}{2\tau_0}) \sin(\omega_0 t)$, where $\omega_0$ is the laser frequency, $a = eE_L/m_e \omega_0 c$ is the dimensionless laser electric amplitude. Here $e$, $m_e$ are fundamental charge and electron mass, $E_L$ is the laser electric field, $c$ is the speed of light in vacuum, respectively. The pulse duration and spot size are defined as $\tau_0$ and $\sigma_0$. A single carbon tube is placed $10\lambda_0$ from the left boundary, with a diameter of $6\lambda_0$. The electron density of the tube when fully ionized is $n_e = 300 n_c$ and the thickness of the wall is $\lambda_0$. The whole target is cold and pre-ionized. To see the full evolvement of a laser pulse in the tube, we set the tube length to be sufficiently long. The cell size is $0.02\lambda_0 \times 0.1\lambda_0 \times 0.1\lambda_0$ to resolve the fine structure and the time step is $\Delta t = 0.008 T_0$ to suppress the numerical instability for high plasma density

When simulating at intensities as high as $10^{23} W/cm^2$, the recoil force on a radiating electron, i.e., the radiation reaction (RR) force must be included in LPI[43-47]. In that case, the code VLPL employs a quantum-electrodynamic (QED) model to calculate the emitted photons and the RR force[48-50]. In our simulations, we turned on the QED model when $a_0 \geq 100$.

**References**

1. Liu, X. *et al.* Mid-infrared optical parametric amplifier using silicon nanophotonic waveguides. *Nat. Photon*. **4**, 557 (2010).
2. Rairoux, P. *et al*. Remote sensing of the atmosphere using ultrashort laser pulses. *Appl. Phys. B* **71**, 573 (2000).
3. Corkum, P. B. Plasma perspective on strong-field multiphoton ionization. *Phys. Rev. Lett*. **71**, 1994 (1993).
4. Noda, S. *et al.* Full three-dimensional photonic bandgap crystals at near-infrared wavelengths. *Science* **289**, 604 (2000).

**Acknowledgements**

This work is supported by the AFOSR Basic Research Initiative (BRI) under contract FA9550-

14-1-0085 and allocations of computing time from the Ohio Supercomputing Center. A. P. is supported by DFG Trnsregio TR18 (Germany).

**Author contributions**

L. L. J. and K. U. A wrote the paper with contributions from J. S., L. L. J. conducted simulations and analysis, J. S. contributed to the proton acceleration part. A. P. developed the code used for simulations (VLPL) and along with R. R. F provided useful suggestions. K. U. A supervised the work,

**Additional information**

Competing financial interests: The authors declare no competing financial interests.